# Inkjet Printed 2D-Crystal Based Strain Gauges on Paper

C. Casiraghi[1], M. Macucci[2], K. Parvez[1], R. Worsley[1], Y. Shin[1], F. Bronte[2], C. Borri[3], M. Paggi[3], G. Fiori[2,*]

[1]School of Chemistry, University of Manchester, UK
[2]Dipartimento di Ingegneria dell'Informazione, Università di Pisa, Pisa, Italy
[3]IMT School for Advanced Studies Lucca, Piazza San Francesco 19, 55100 Lucca, Italy

*Corresponding author email:gfiori@mercurio.iet.unipi.it

**Abstract**

We present an investigation of inkjet printed strain gauges based on two-dimensional (2D) materials. The technology leverages water-based and biocompatible inks to fabricate strain measurement devices on flexible substrates such as paper. We demonstrate that the device performance and sensitivity are strongly dependent on the printing parameter (i.e., drop-spacing, number of printing passes, etc.). We show that values of the Gauge Factor up to 125 can be obtained, with large sensitivity (>20%) even when small strains (0.3%) are applied. Furthermore, we provide preliminary examples of heterostructure-based strain sensors, enabled by the inkjet printing technology.

## 1. Introduction

Strain sensor technology is playing a vital role in a wide range of applications, spanning from the detection of subtle and large human motions [1][2], the reproduction of human skin sensory ability in robotics and health monitoring applications [3][4][5][6][7], and screening the reliability and durability of automotive, marine and aerospace structural systems [8][9].

Conventional strain gauges, while able to provide rich information about the mechanical stress, are typically cumbersome, costly to install and maintain (particularly for large-scale structures), they often require complex signal processing schemes and their replacement can be difficult. It is clear that such limitations can be prohibitive especially when such systems are used in wearable and portable electronics devices.

As a consequence, low-cost, flexible, and easy to integrate strain gauges are needed in order to overcome these problems. From this point of view, printed strain sensors on paper could match the aforementioned requirements, due to the flexibility and extremely low cost [10] of such a solution.

Piezoresistive sensors, which translate strain ($\varepsilon$) variations into changes of resistance ($R$) are widely used, due to their simple design and readout mechanism [11]. One of the most important figures of merit is the Gauge Factor ($GF$), which describes the change in resistance ($\Delta R=R-R_0$, where $R_0$ is the nominal resistance) due to mechanical deformation, and can be expressed as $GF= (\Delta R/R_0)/\varepsilon =S/\varepsilon$, where $S$ is the sensitivity. Note that metal foil gauges generally attain $GF$ values of 2–4, while screen-printed polymer thick film sensors attain $GF$ values ranging from 8 to 14 [12]. Semiconducting foil strain sensors can perform better, but they are fragile and work only with limited strain values [13].

Graphene-based strain sensors have attracted much attention because of the unique properties of this novel material, such as very high conductivity, mechanical strength, and flexibility [14][15]. Most of graphene-based piezoelectric sensors have been fabricated with graphene produced by means of Chemical-Vapour Deposition (CVD) [14][16]. However, this method requires expensive vacuum systems and time consuming post-growth processing, such as etching and transferring onto the flexible substrate. The transfer process can introduce damage on the graphene and leave polymer residuals, which can affect the performance of the strain gauge [17] and limits the reproducibility of this fabrication method. Furthermore, the $GF$ predicted for graphene is rather low (~2.4 from calculations in [18]), due to its semi-metal character. Strategies such as cutting graphene into ribbons or introducing wrinkles have been used to increase the *GF* [17][19], further complicating the fabrication of such sensors and reducing their reliability.

The use of solution-processed graphene offers a simple and low-cost way to easily fabricate the devices. However, in most cases, graphene is used as filler in a polymer or elastomer matrix or with carbon nanotubes (CNTs) to form a composite [20][21][22][23][24][25]. Therefore, the fabrication of the sensor requires several chemical processing steps; furthermore, most sensors still need to be glued on the surface to be inspected: typically, this is done by using epoxy glue [26], which requires high temperature curing (at about 80-100 °C) and careful preparation of the substrate surface. Ideally, one would like to fabricate the sensor directly on the surface to be inspected, possibly on curved or complex surfaces, without using any surface or sintering treatment, in order to integrate the device into a wide range of flexible substrates, such as paper.

To the best of our knowledge, only very few studies report the direct fabrication on the substrate of graphene-only strain gauges (Table 1). Values of the *GF* between 4 and 200 have been obtained (Table 1), depending on the material and on the method used to fabricate the device. Spray coating and screen printing techniques have been previously used [13][25][27]. However, these methods typically produce relatively thick (and therefore less flexible) films, and they are characterized by a large amount of waste material and limited resolution and design flexibility (e.g. a shadow mask is required to fabricate the device). Alternatively, a layer-by-layer assembly has been used [2], which requires several chemical processing steps and has limited design flexibility. Self-assembly [28] requires transfer of the film from the liquid to the flexible substrate. Finally, many of those approaches involve handling of toxic solvents and chemicals, and none of the strain gauges were fabricated directly on paper (Table 1). Only one seminal paper [29] reports a graphite-based strain gauge on paper made with a simple pencil. Therefore, there is the need to investigate low cost and mass scalable techniques, requiring minimal chemical processing, for direct fabrication on paper of graphene-based strain gauges.

In this study, we have fabricated graphene strain gauges on paper by means of inkjet printing, with an extensive investigation over the printing parameters (i.e., print passes, drop-spacing). Such an investigation is also of primary importance towards a complete understanding of the electrical properties of conductive graphene lines on flexible substrates, one of the main building blocks for printable electronics based on 2D materials, and their combinations through lateral and vertical heterostructures. Water-based and biocompatible inks of graphene were used with no pre- or post-processing of the devices [30]. A maximum *GF* of 125 has been obtained, associated to large sensitivity (>20%) even when small strains

(0.3%) are applied. The advantage of inkjet printing is in its design flexibility, in particular in the possibility to build devices of arbitrary geometry in both the planar and the vertical directions. In this respect, we also shown preliminary results on heterostructure based strain gauges, made of graphene and hexagonal Boron Nitride (hBN). These initial devices show a higher *GF*, although more studies are needed to optimize the fabrication process and device performance.

## 2. Experimental

### 2.1 Material Preparation

The graphene ink is formulated from graphite via ultrasonic-assisted liquid phase exfoliation [31] in water [30][32][33][34]. In details, 1.5 g of graphite flakes (Sigma Aldrich, 100+ mesh) and 500 mg of 1-Pyrenesulfonic acid sodium salt (PS1, from Sigma Aldrich) were mixed into 500 mL of de-ionized (DI) water. The mixture was then sonicated at 300W using a Hilsonic bath sonicator for 5 days. Afterwards, unexfoliated graphite was removed by centrifugation (Sigma 1-14k refrigerated centrifuge) at 4000 rpm (1180g) for 20 minutes. The supernatant containing graphene and PS1 in water was collected and then centrifuged again at 15000 rpm for 1 h to collect the sediment. After centrifugation, the supernatant containing excess amount of PS1 in water was discarded. The precipitate was re-dispersed in the printing solvent, whose composition is described in [30]. The same process is used for h-BN (Sigma Aldrich, 98% purity, 1 μm size particles).

### 2.2 Characterization

The final concentration of graphene is determined by using the Beer-Lambert law and an absorption coefficient of 2460 L $g^{-1}$ $m^{-1}$, measured at 660 nm for graphene [35][36][37] and an absorption coefficient of 1000 L $g^{-1}$ $m^{-1}$ measured at 550 nm for h-BN [38]. A Perkin-Elmer l-900 UV-Vis-NIR spectrophotometer was used to acquire the spectra.

A Bruker Atomic Force Microscope (MultiMode 8) in Peak Force Tapping mode, equipped with ScanAsyst-Air tips is used to determine the lateral size distribution of the flakes. The sample was prepared by drop casting the solution on a clean silicon substrate; several areas of 100 $\mu m^2$ were scanned and about 130 flakes were selected for lateral size analysis. The same sample preparation has been used for Raman measurements; about 50 isolated flakes were measured. Raman measurements were performed using a Renishaw Invia Raman spectrometer equipped with a 514.5 nm excitation line with 1.0 mW laser power. 100X NA0.85 objective lens and 2400 grooves/mm grating were used.

A Dimatix DMP-2850 inkjet printer (Fujifilm Dimatix, Inc., Santa Clara, USA) was used. This equipment can create and define patterns over an area of about 200 mm x 300 mm and handle substrates that are up to 25 mm thick. A waveform editor and a drop-watch camera system allows manipulation of the electronic pulses to the piezo jetting device for optimization of the drop characteristics as it is ejected from the nozzle. The nozzle plate consists of a single row of 16 nozzles with a 23 µm diameter, spaced 254 µm apart, with typical drop size of 10 pl. Among the many types of papers available on the market, we have selected the PEL P60 paper (from Printed Electronics Limited), characterized by a micro-porous surface treatment, designed to wick away the carrier solvent of the ink, while allowing for uniform deposition.

Electrical measurements have been performed by means of a Keithley 4200-SCS Parameters Analyzer at room temperature and in air, in order to evaluate the variation of the resistance of the strain sensor at different applied strain values. Strain has been imposed by constraining the strain gauge on substrates with different curvatures. As in [27], the applied strain ($\varepsilon$) can be expressed as $\varepsilon = t/2r$, where $t$ is the thickness of the paper substrate (250 µm) and $r$ the radius of the substrate on which the strain gauge is deposited (in our case, this is between 4.4 and 1 cm, leading to a maximum strain of 1.25 % with our setup).

3. **Results**

Figure 1a shows the UV-Vis spectrum of the graphene ink (diluted 100x) used to print the strain gauges. Using the Beer-Lambert law, we estimated a concentration of ~1.81 mg/mL, which is high enough to print conductive graphene lines with few printing passes [30].

Figure 1b shows the distribution in lateral size of the nanosheets: most of the flakes have lateral size between 50 and 400 nm. The majority of the material has a lateral size of ~200 nm, which satisfies the inkjet printer requirements to avoid nozzle blockage.

Figure 1c shows some representative Raman spectra measured on individual flakes. Raman spectroscopy is a very powerful technique for the characterization of graphene [39][40]. The typical Raman spectrum of pristine graphene shows the D and G peaks, placed at about 1350 and 1580 cm$^{-1}$, respectively [41]. A single and sharp 2D peak is typically used to identify graphene [41]. However, the Raman spectrum of liquid-phase exfoliated graphene is strongly affected by the exfoliation process: during sonication, the nanosheets are subjected to strong mechanical stress originating from the process of formation and collapse of bubbles and voids in the liquid. This ultimately breaks the flakes into smaller and thinner pieces. This is reflected in the Raman spectrum, which is typically characterized by the D peak – this mode is activated by the edges of the nanosheets, having typical size comparable to or smaller than that of the laser spot [42]. Another effect is observed on the 2D peak, which can show complex lineshapes, likely due to folding and re-stacking of the flakes. In previous studies [43][44][45][46], we introduced and tested a simple qualitative method based on the shape and symmetry of the 2D peak to distinguish between single-layer graphene, few-layer sheets and graphitic material (> 10 layers with AB stacking, intended as residual graphite). Using this method, we found that 20-30% of the flakes are single-layer, while the majority of the flakes are few-layers.

In the inset of Figure 2a, we show the layout of the printed graphene strain sensor. The graphene line is 10 mm long and 0.5 mm wide. Two graphene pads (10 x 5 mm$^2$) have been also printed, to act as contacts. We have indeed experienced that using silver paste to directly connect the graphene line ends to wires does lead to graphene delamination, due to the stiffness of the paste. In order to solve such an issue, we have printed graphene pads, to which crocodile clips have been applied on top of copper foils, in order to avoid direct contact of the clips to the sample (to prevent sample damage). Alternatively, we have observed that printed graphene pads can be substituted with a conductive glue (from Bare conductive), which assures high conductivity and excellent mechanical properties.

Figure 2a shows the resistance of a graphene line made with 10 print passes as a function of the number of bending cycles (with a maximum strain of 1.25%). Here, and in the following experiments, the bending has been applied along the length of the strain sensor, parallel to the direction of current flow. The graphene line resistance remains almost constant through the bending cycles, with a maximum variation of 0.46% with respect to the mean value (i.e., 266.78 kΩ). In Figure 2b, we show the sensitivity as a function of time, while applying a positive and a negative strain, which leads to an increase (positive strain) and a decrease (negative strain) of the nominal resistance.

In order to engineer the strain sensor, we have investigated the electrical behaviour of the graphene lines under different strains and for different printing parameters, as for example the number of printing passes and the drop spacing (unless otherwise specified, we refer to a drop spacing of 20 μm, which has been found to be optimal for minimizing the sheet resistance of the printed graphene lines on PEL P60 paper [30]). In particular, in Figure 3a, we show the resistance as a function of the inverse of the bending radius (1/$r$), and for different numbers of layers. As expected [30], the larger the number of printing passes (i.e., printed layers), the smaller the resistance. We have also considered a serpentine sensor (as the one shown in the inset of Figure 3a) made with 15 print passes. In this case, the resistance is larger in comparison with that of the other device, due to the reduced number of layers and the larger effective length. We have then extracted the sensitivity as a function of the applied strain, as shown in Figure 3b. As can be seen, we observe a larger *S* for the sample with the larger number of layers, reaching a sensitivity greater than 100 %, for a curvature of 1 cm, i.e., a strain of 1.25%. For negative strain, instead, *S* seems to be independent of the strain sensor thickness. Note that the *GF* anisotropy between tension and compression was also observed with graphite pencil on paper [29]. Among the considered devices, the one with the serpentine shape shows the smallest sensitivity (grey squares in figure 3b), since, while applying the strain along the longitudinal direction, the line segments printed in the normal direction do not change their resistance, but still contribute to the overall resistance, eventually leading to a reduction of *S*.

The increased sensitivity with the larger number of layers is in contrast with previous results presented in [13] and [47], where the opposite trend has been observed, i.e., larger *S* and larger gauge factors have been observed for decreasing concentration/thickness of the material. Such behaviour has been attributed to percolating path transport in the graphene network. In order to investigate this problem, we have printed strain devices using larger drop-spacing: this is expected to decrease the uniformity of the line, and therefore the probability of the flakes to be in contact or overlapped over a large area.

Figure 4a shows the *GF* as a function of the resistance of the strain sensor, while considering two different values of the drop spacing, i.e., 40 and 70 µm. As can be seen, in this case, results are qualitatively in agreement with Ref. [13], i.e. the gauge factor increases for decreasing graphene resistance. This points out that tuning the printing parameters allows manufacturing strain sensors with very different characteristics. This also indicates that the *GF* strongly depends on how the flake concentration is distributed on the substrate.

To shed a light on this effect, we have performed a mechanical and morphological investigation of the strain gauge, through the exploitation of two different measurement systems: a confocal profilometer with a lateral resolution of 1.66 µm and a scanning electron microscope (SEM) for in-situ micromechanical testing and inspection of microcracks nucleation and propagation. Results are shown in the Supplementary Material. As can be seen in Fig. S1, small drop spacing and thicker films can induce microcracks with higher density and larger crack openings upon strain. This higher sensitivity to deformation leads eventually to a higher gauge factor, as also observed in [49] in reduced graphene oxide.

As already observed in [30], the porosity of the paper-based substrate allows printing with very small drop spacing, leading to deposition of a larger amount of material per unit area with fewer printing passes. From an engineering point of view, such an observation is very important, since it introduces new degrees of freedom for the design of strain sensors. In the case of traditional strain sensors, the *GF* is constant, while, by inkjet printing graphene lines, it becomes possible to tune the electro-mechanical properties of the sensor, acting upon the strain resistance and the printing parameters, and to tailor the sensor to specific needs and applications.

One of the main advantages of inkjet printing compared to other deposition techniques is the possibility to fabricate complex devices, such as arrays of sensors or vertical heterostructures [30]. Here we exploit this flexibility in design, by showing the first heterostructure-based strain gauge, made of graphene and hexagonal Boron Nitride (hBN). The inset of Figure 4b shows the schematic of the device, consisting of a heterostructure made of hBN at the bottom and graphene on top (hBN/Gr). Figure 4b compares the *GF* of the hBN/Gr strain gauge (blue diamonds) with those obtained with graphene-only strain gauges, printed with 20 µm drop spacing. Such devices show the same qualitative trend as in the case of the graphene-only strain gauge (i.e. larger *GF* for smaller resistance), but have larger *GF* for the same values of resistance. This is qualitatively in agreement with [50], where the presence of a layer between the (rubber) substrate and the graphene sensor led to an increase of the *GF* for the same value of resistance. The hBN layer may decrease the roughness of the paper and improve the adhesion of the device on the substrate. This opens up the possibility of adding a further degree of freedom to the device design space, towards the objective to achieve high-performance and multi-functional strain sensors, by introducing different 2D materials and more complex geometries.

In Figure 5a, we show the schematic of a simple circuit employing the strain sensor as a variable resistance in series to an LED (embedded on a paper substrate) and an external battery. As shown in Figure 5b, when tensile strain is applied, the resistance of the strain sensor increases, thus reducing the current and the luminosity of the LED. On the other hand, when compressive strain is applied, the resistance decreases, and the current increases, as well as the LED brightness. This simple circuit could find application as a first warning of an anomalous strain condition, in particular combining it with an energy scavenger for the power supply (e.g. to replace the battery).

## 4. Conclusions

We have demonstrated inkjet printed graphene-only strain gauges on paper with a gauge factor close to 150. Inkjet printing allows simple and fast fabrication of the sensor directly on the surface to be inspected, opening the possibility to define arrays of sensors over large areas or multisensing, by introducing different types of sensors in the array. Finally, inkjet printing allows full flexibility in the design of strain gauges, with the exploration of the printing parameters (i.e., drop-spacing, number of print passes etc.), and also with the possibility of exploiting combinations of different 2D materials (e.g., graphene and hBN). We have also shown some preliminary results on heterostructure-based strain gauges, which may inspire, after further investigation and optimization, new concepts in the space of multi-functional strain gauges made of 2D materials.

## 5. Acknowledgements

CC and KP acknowledge the Grand Challenge EPSRC grant EP/N010345/1. YS acknowledges the EPSRC project "Graphene-Based Membranes" (EP/K016946/1). RW acknowledges the Hewlett-Packard Company for financial support in the framework of the Graphene NowNano Doctoral Training Center. GF gratefully acknowledgesthe Project 'Graphene Flagship' Core 1 (contract no. 696656). GF and MP acknowledge financial support from Fondazione Cassa di Risparmio di Lucca.

films with layered structure, Composites Part A: Applied Science and Manufacturing 80 (2016) 95 – 103.

**Table 1** Summary of the publications reporting direct fabrication of strain gauges based only on graphene, using solution processing based methods. GNPs= graphene nanoplatelets; PSS= polystyrene sulfonate; PDMS= Polydimethylsiloxane; NMP= N-Methyl-2- pyrrolidone; PET= Polyethylene terephthalate; PEN= Polyethylene naphthalate; CNT= carbon nanotube; GO= Graphene Oxide.

| Reference | GF | Material | Fabrication | Substrate |
|---|---|---|---|---|
| [2] | 9-1100 at 5% strain (no patterned substrate) | Commercial GNPs in water, mixed with PSS | Layer-by-layer with polymers | Transfer on PDMS |
| [13] | 10-200 (max strain: 2%) | Microwave exfoliation, followed by NMP dispersion and solvent exchange | spray coating | PET |
| [25] | 7.8-4 (for increasing CNTs amount) at 0.2% strain | CNT(bottom)/GNP(top) hybrid thin films; aqueous dispersions prepared by sonication with surfactant | spray coating, 90 °C | PET |
| [27] | 19.3 at 0.7% strain | Commercial graphite ink | screen printing + thermal curing (120 °C, 30 mins) | PEN |
| [28] | 500 at 1% strain | Electrochemical exfoliation, followed by dispersion in ethanol | Self-assembly technique | Transfer on PDMS |
| [48] | 0.11 (film) 9.49 (ribbons) | Reduced GO films and ribbons (20 µm width) | laser scribing used for reduction | PET |
| [50] | Max 35 (max strain: 5%) | GNPs in water, mixed with surfactant | Spray coating 100 °C | Rubber, covered with polymer film |
| Our work | Max 125 | Water-based graphene inks | Inkjet-printing | paper |

**Figure 1 a)** UV-Vis spectrum of the graphene ink used (diluted 100X); inset: picture of the ink. **b)** Lateral size distribution of the flakes as measured by AFM. **c)** Representative Raman spectra measured on isolated flakes.

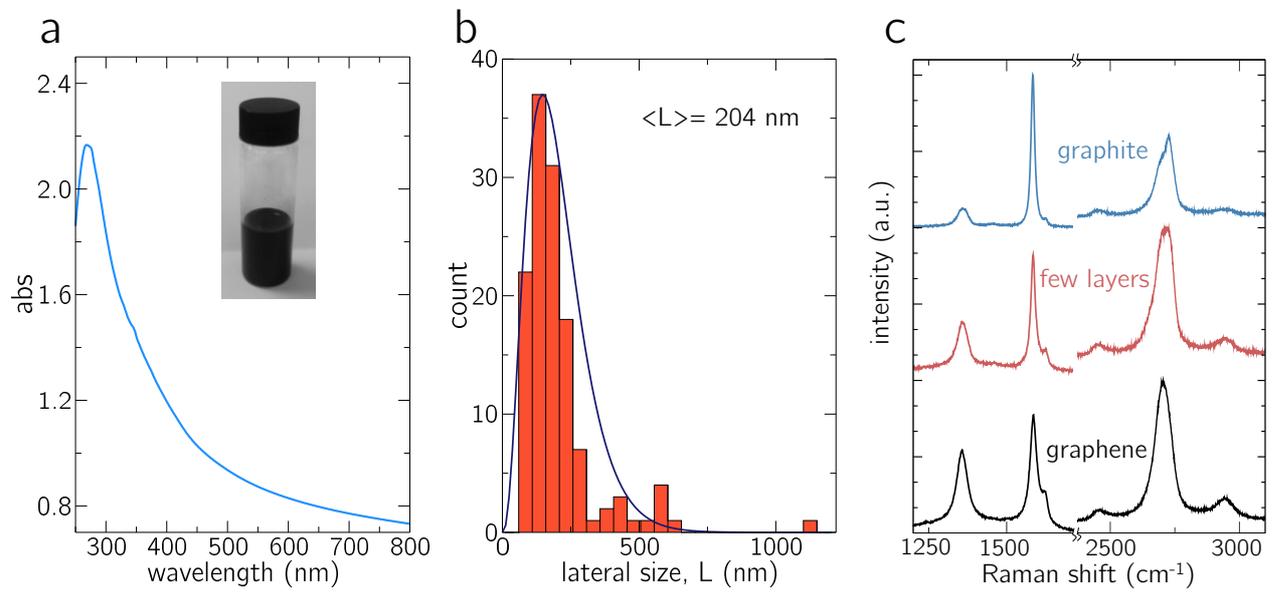

**Figure 2 a)** Resistance of a 10 layers graphene line as a function of the number of bending cycles, for a maximum applied strain of 1.25%. In the inset, the layout of the printed graphene strain sensor. **b)** Sensitivity as a function of time, while applying a positive and a negative strain.

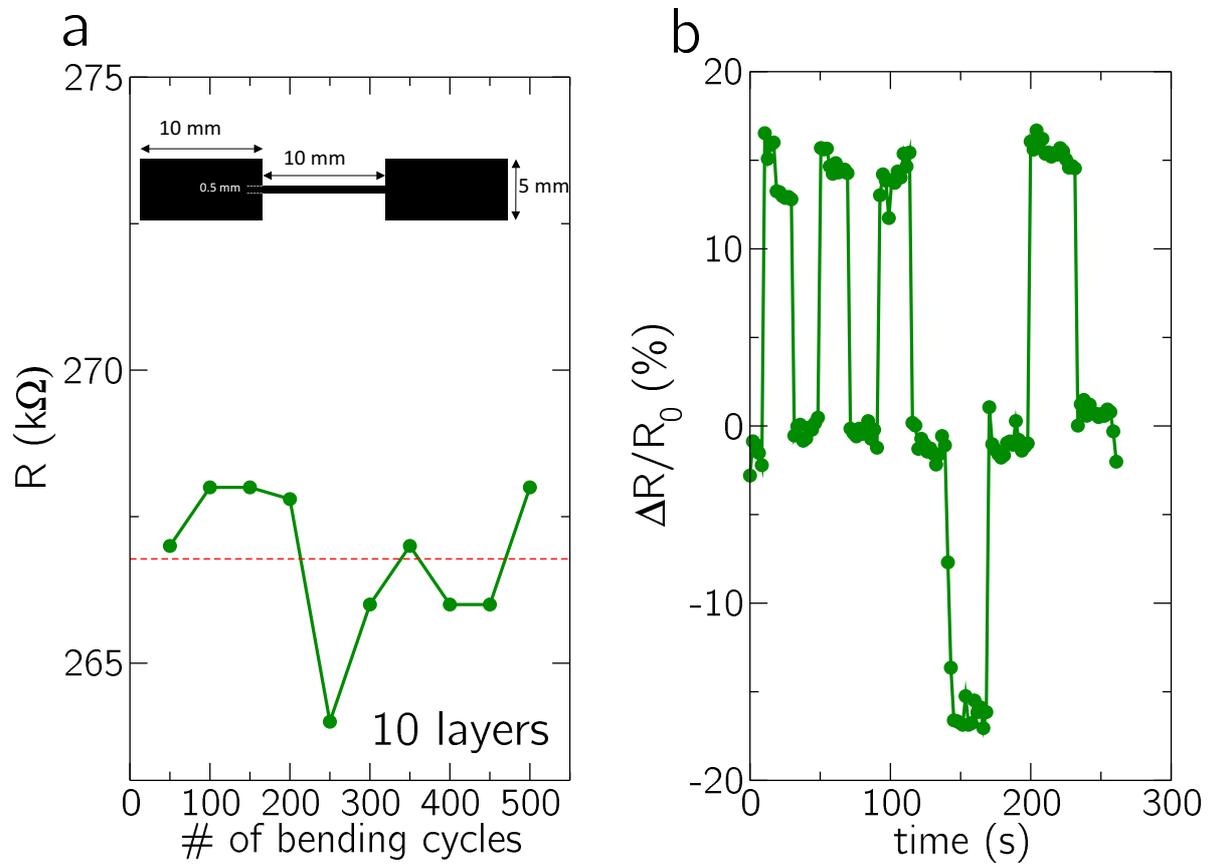

**Figure 3 a) Resistance as a function of the inverse of the bending radius and for different layer thicknesses. All the results refer to devices with the layout shown in the inset of Figure 2a, except for the grey (square symbol) curve, which refers to the serpentine layout shown in the inset. b) Sensitivity as a function of 1/*r* for different layers.**

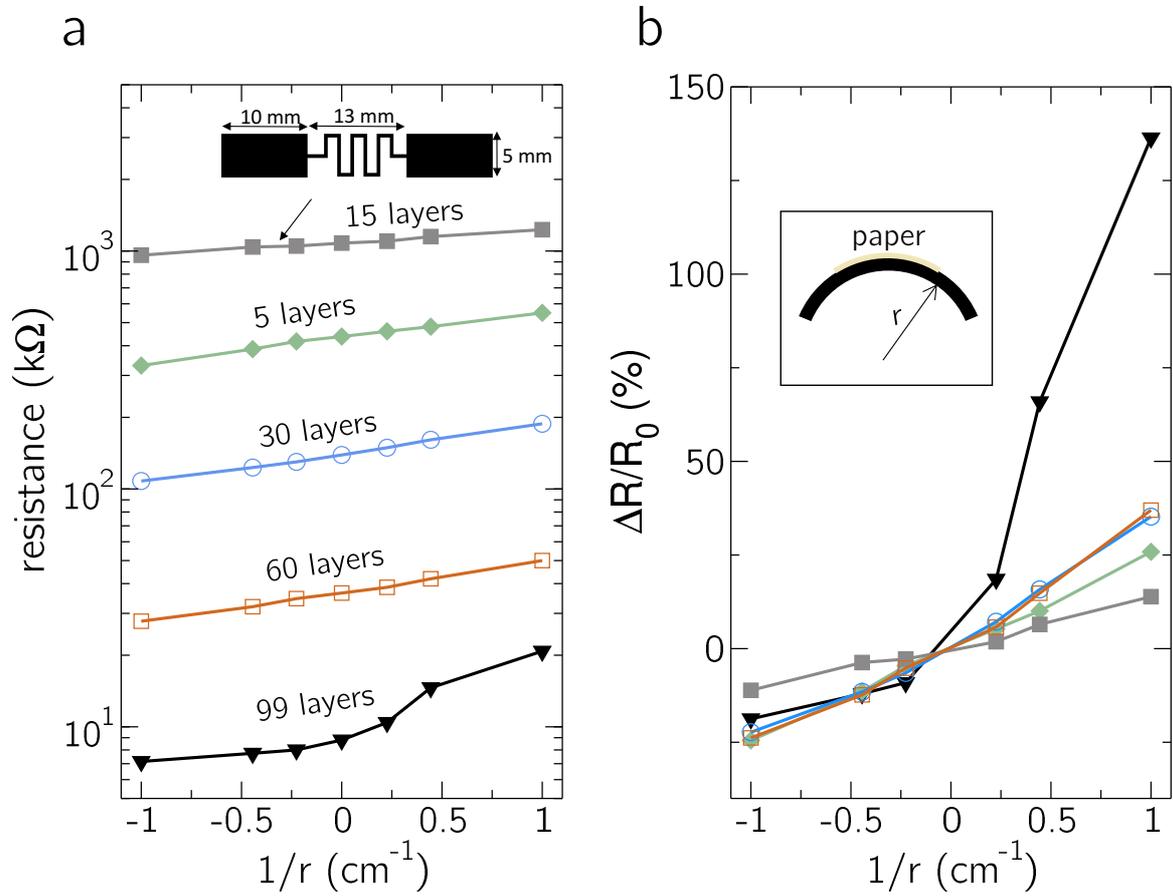

Figure 4 a) Gauge Factor as a function of strain sensor resistance for two different drop-spacings, i.e., 40 and 70 μm. b) Gauge Factor as a function of the resistance, for a drop-spacing of 20 μm. In both figures, the dashed lines are a guide for the eye. Blue dots correspond to graphene/hBN heterostructures.

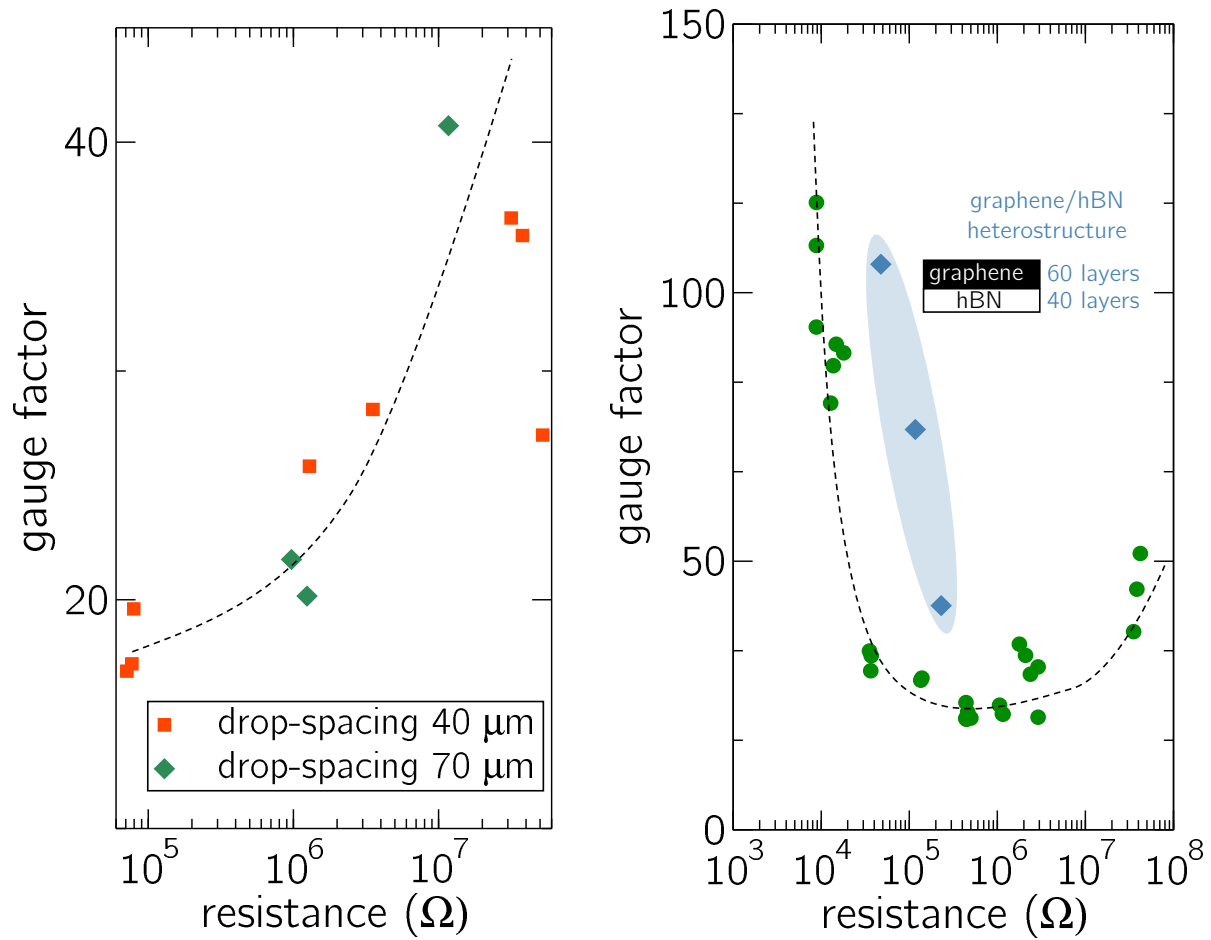

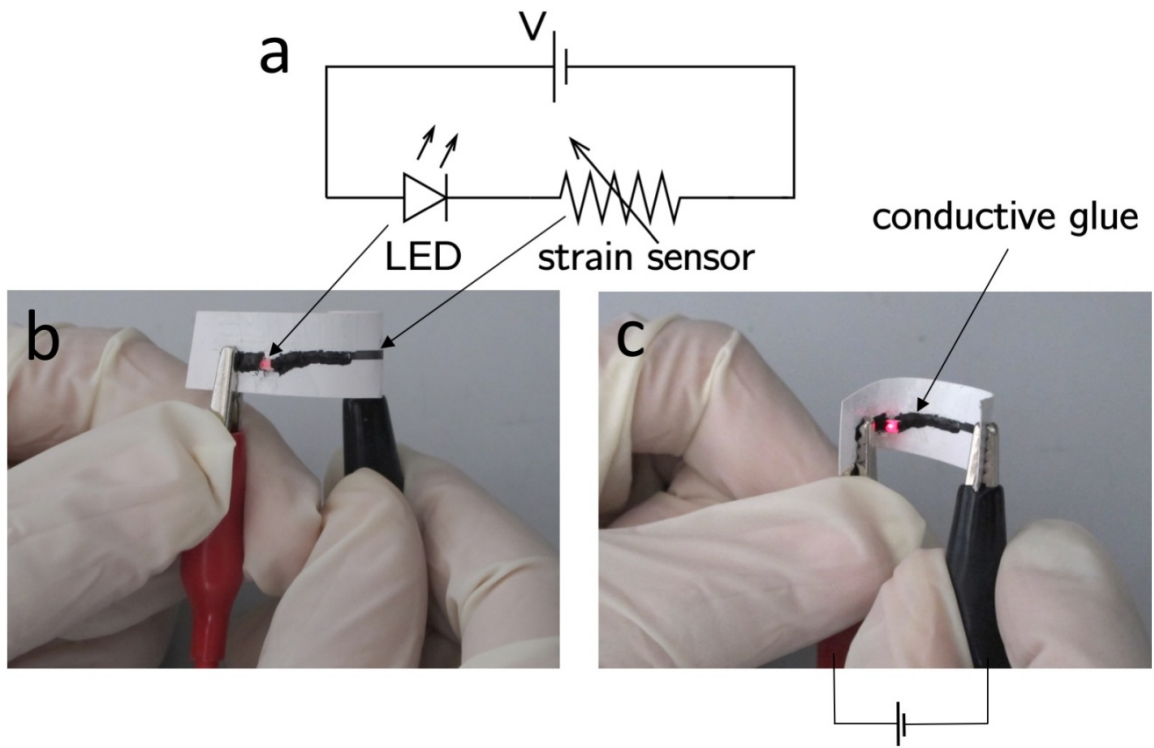

Figure 5 a) Schematic of the simple circuit fabricated on a paper substrate and employing the graphene strain sensor. System under b) tensile and c) compressive strain.